\documentclass[11pt,a4paper]{article}
\pdfoutput=1
\usepackage{jheppub}
\usepackage[utf8]{inputenc}
\usepackage{graphicx}
\usepackage{amsmath}
\usepackage{amsfonts}
\usepackage{mathrsfs}
\usepackage{amssymb}
\usepackage{subfigure}
\usepackage{multicol}
\usepackage{epsfig}
\usepackage{epstopdf}
\usepackage{ulem}
\usepackage{array}
\usepackage{slashed}
\usepackage{parskip}

\newcommand{\GeV}{{\rm GeV}}
\newcommand{\TeV}{{\rm TeV}}

\newcommand{\blue}[1]{\color{blue} #1 \color{black}}

\newcommand{\pmatr}[1]{\begin{pmatrix} #1 \end{pmatrix}}

\graphicspath{{figures/}}

\title{\blue{Constraints from Triple Gauge Couplings on Vectorlike Leptons}}

\author[a]{Enrico Bertuzzo,}
\author[b]{Pedro A. N. Machado,}
\author[a]{Yuber F. Perez--Gonzalez,}
\author[a]{Renata Zukanovich Funchal}

\emailAdd{bertuzzo@if.usp.br}
\emailAdd{pmachado@fnal.gov}
\emailAdd{yfperezg@if.usp.br}
\emailAdd{zukanov@if.usp.br}

\affiliation[a]{Departamento de F\'isica Matem\'atica, Instituto de F\'isica, Universidade de S\~ao Paulo,\\ 
Rua do Mat\~ao 1371, CEP.\ 05508-090, S\~ao Paulo, Brazil}
\affiliation[b]{Fermi National Accelerator Laboratory, Batavia, IL, 60510, USA}

\abstract{We study the contributions of colorless vectorlike fermions 
  to the triple gauge couplings $W^+ W^- \gamma$ and $W^+ W^- Z^0$. We consider  
  models in which their coupling to the Standard Model Higgs boson is allowed or 
  forbidden by quantum numbers.  We assess the sensitivity of the  future 
  accelerators FCC-ee, ILC and CLIC to the parameters of these models, assuming 
  they will be able to constrain the anomalous triple gauge couplings with a 
  precision $\delta \kappa_V \sim \mathcal{O}(10^{-4})$, $V=\gamma,Z^0$. We show 
  that the combination of measurements at different center-of-mass energies helps 
  to improve the sensitivity to the contribution of vectorlike fermions, in 
  particular when they couple to the Higgs. In fact, the measurements at the FCC-ee and, 
  especially, the ILC and the CLIC, may turn the triple gauge couplings into a new 
  set of precision parameters able to constrain the models better than the oblique 
  parameters or the $H \to \gamma \gamma$ decay, even assuming the considerable improvement 
  of the latter measurements achievable at the new machines.\\

  \begin{center}
     Published in {\it Phys.\ Rev.\ D}  
  \end{center}   
  
  }
  
\keywords{}

\begin{document}
\begin{flushright}
	FERMILAB-PUB-17-190-T
\end{flushright}

\maketitle

\section{Introduction}\label{sec:intro}

All experimental data collected so far have confirmed the Standard Model (SM) predictions, 
including the existence of a scalar particle that seems to have the right properties 
to match those of a Higgs boson. The SM cannot, however, be the final theory of particle 
physics, since it does not explain neutrino masses nor the baryon asymmetry of the Universe 
and it does not contain a dark matter (DM) candidate. Moreover, if the naturalness principle 
applies, new physics (NP) is expected. 

The nature of the NP models that are supposed to complete the SM is
elusive and unknown. Taking a bottom-up approach, however, we can
suppose that, exactly as the SM particles are vectorlike from the low-energy 
QED/QCD point of view, the first particles to be discovered (if
any) will be vectorlike from the SM point of view \cite{Kilic:2009mi}. 
In addition, vectorlike fermions arise in
many well-motivated SM extensions such as models with extra
dimensions~\cite{Frampton:1999xi,Agashe:2008fe,Gopalakrishna:2013hua,Fichet:2013ola},
composite Higgs~\cite{Kaplan:1991dc,Redi:2013pga,Contino:2004vy}, two-Higgs-doublet-model 
extensions~\cite{Dermisek:2015oja}, low-scale
supersymmetry~\cite{Martin:2009bg,Graham:2009gy} and, more recently,
in new solutions of the hierarchy problem \cite{Graham:2015cka,Arvanitaki:2016xds}.  
Vectorlike fermions are much less constrained than extra chiral families, 
which in fact are now pretty much ruled out by data after the observation of the 125
GeV boson at the LHC~\cite{Anastasiou:2011qw,Anastasiou:2016cez}. Vectorlike quarks
masses are typically bounded from ATLAS and CMS Run 1 data to be $\gtrsim$
(800--1000) GeV \cite{Chatrchyan:2013uxa,Chatrchyan:2013wfa,Khachatryan:2015axa,
Khachatryan:2015gza,Aad:2015mba,Aad:2015kqa,Aad:2015voa,Aad:2016qpo}, while direct
constraints on vectorlike leptons come only from the LEP experiments and are constrained to be $\gtrsim$ 100
GeV \cite{Achard:2001qw}. Bounds from  electric and magnetic dipole moments 
and electroweak precision measurements have been also considered \cite{Altmannshofer:2013zba,Cai:2016sjz}.

As no new particles have been discovered so far, there is growing
interest in the community in future $e^+$ $e^-$ colliders that could
pursue the electroweak precision tests started by LEP and the SLC
profiting of higher energies and luminosities. This moves from the
observation that, for heavy enough particles, NP may first show up
through loop effects, and as such be bounded by electroweak precision
measurements, modifications of $H\to \gamma\gamma$ or anomalous triple
gauge couplings (TGCs). In particular, the new machines can probe the
anomalous TGCs $W^+W^-\gamma$, $W^+W^-Z^0$, and $Z^0Z^0\gamma$ to
unprecedented levels. Since the structure of the TGCs is a direct
manifestation of the non-Abelian nature of the SM gauge group, they
are sensitive to the presence of NP with $SU(2)_L \times U(1)_Y$
representation and, in particular, to the presence of vectorlike
fermions.

The purpose of this paper is to estimate the sensitivity of future $e^+$
$e^-$ machines to vectorlike leptons, in many  possible realizations, via the
measurements of triple gauge couplings which will putatively reach  a ${\cal O}(10^{-4})$ precision.
The paper is organized as follows. In
Sec.~\ref{sec:TGC} we start by defining the TGCs form factors that
can be modified by SM loop corrections and new physics. Next, in
Sec.~\ref{sec:VLF} we describe the vectorlike lepton models that we
will study in this paper and how they can contribute to the TGCs form
factors.  In Sec.~\ref{sec:Constraints} we estimate the constraints on
these models that can be achieved by TGCs measurements at three
proposed future accelerator facilities: the Future Circular Collider
(FCC-ee)~\cite{dEnterria:2016sca}, International Linear Collider
(ILC)~\cite{Baer:2013cma}, and the Compact Linear Collider
(CLIC)~\cite{Dannheim:2012rn}.  Finally, in Sec.\ref{sec:conc}, we
discuss our conclusions.

\section{Triple gauge couplings}
\label{sec:TGC}
%
\begin{figure}[tb]
  \centering
	\includegraphics[scale=0.35]{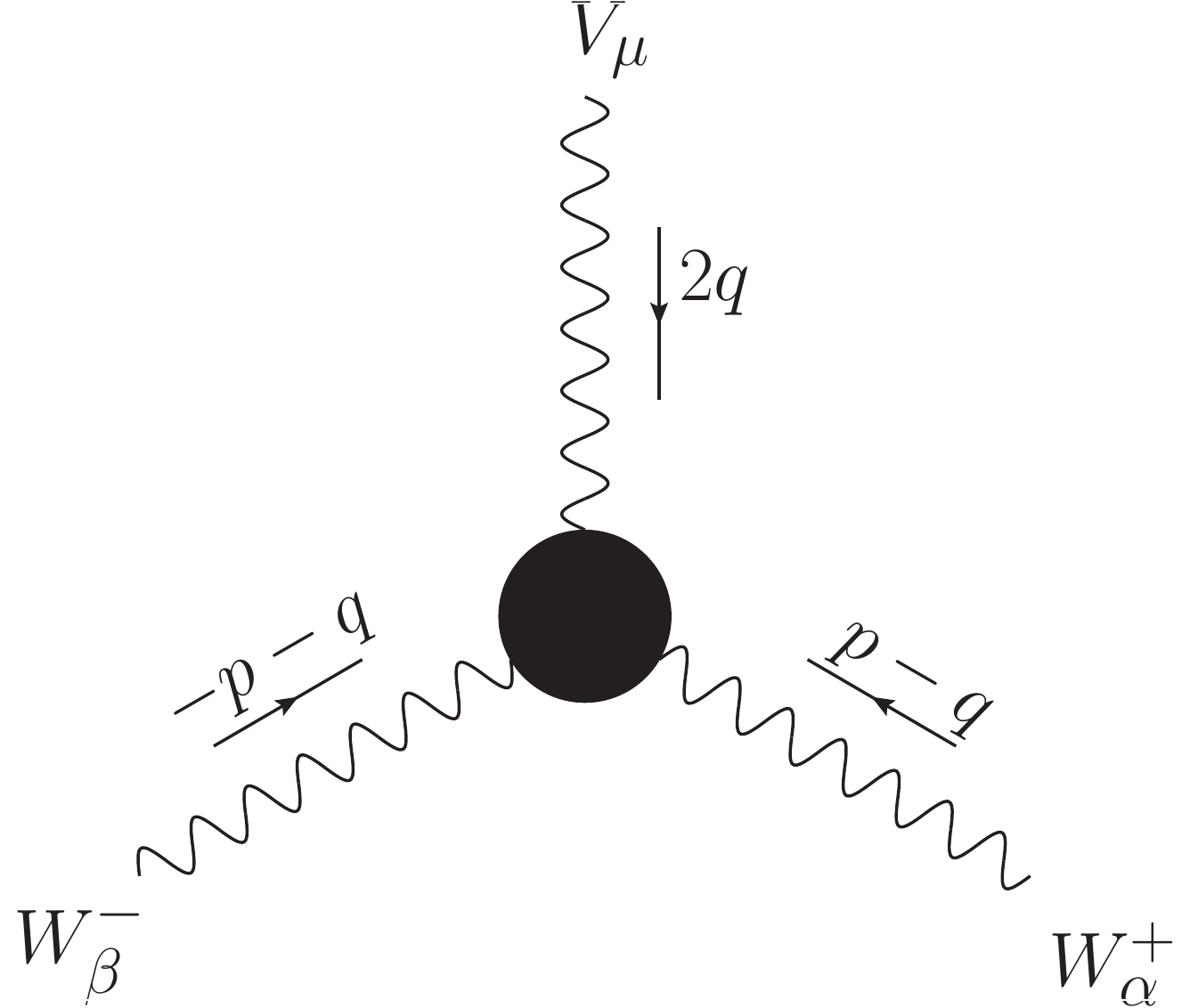}
	\caption{Feynman diagram of the $WWV$ vertex in momentum space.}
	\label{fig:TGC}
\end{figure}
The typical structure of the charged TGCs that we will consider in this paper is shown in Fig.~\ref{fig:TGC}, where $V$ can be either the $Z^0$ boson or the photon. The complete one-loop SM contribution to the charged TGCs $W^+W^-\gamma$ and $W^+W^-Z^0$ was computed some time ago~\cite{Argyres:1992vv,Papavassiliou:1993ex,Fleischer:1991vq}, while the contribution to the neutral TGC $Z^0Z^0\gamma$ was studied in Refs.~\cite{Argyres:1992vv,Gounaris:2000tb}. The charged couplings can be directly studied in future $e^+ e^-$ colliders, through $e^+ e^- \to W^+ W^-$. The neutral couplings, on the other hand, can be studied using the processes $e^+ e^- \to Z^0 \gamma$ or $e^+ e^- \to Z^0 Z^0$, with subsequent decays $Z^0 \to \bar{\nu}\nu$ and $Z^0 \to \ell^+ \ell^-$\cite{Khachatryan:2015kea,Khachatryan:2015pba,Khachatryan:2016yro}. Let us note that only fermions with an axial coupling to the $Z^0$ boson can generate nonvanishing corrections to the neutral TGCs~\cite{Gounaris:2000tb}. As such, since our focus is 
vectorlike fermions, we will just consider the effects on the charged verteces. 

The generic charged TGC vertex $WWV$, with $V=\gamma, Z^0$, can be parametrized using the effective Lagrangian~\cite{Hagiwara:1986vm} 
\begin{align}\label{eq:EfLag}
	\mathcal{L}_{WWV}=&-ig_{V}[(W_{\mu\nu}^\dagger W^\mu V^\nu-W_{\mu\nu} W^{\mu\dagger} V^\nu)+\kappa_V W_\mu^\dagger W_\nu V^{\mu\nu}
	+\frac{\lambda_V}{M_W^2}W_{\mu\tau}^\dagger W^\tau_{\,\,\,\nu} \,V^{\nu\mu}\notag\\
	&+\mathcal{L}_{WWV}^\textrm{nCP},
\end{align}
where $\mathcal{L}_{WWV}^\textrm{nCP}$ contains $P$ or $C$ odd terms, $\kappa_V$ and $\lambda_V$ are form factors, the field strengths are defined as $W_{\mu\nu}=\partial_\mu W_\nu-\partial_\nu W_\mu$~\footnote{Notice that, with this definition, the $W$ field strength is not $U(1)_{em}$ invariant. New quadrilinear terms must be introduced in $\mathcal{L}$ to make the whole Lagrangian gauge invariant.} and
$V_{\mu\nu}=\partial_\mu V_\nu-\partial_\nu V_\mu$, and the coupling $g_V$ is given by
\begin{align}
	g_{V}=\begin{cases}
	 e & \textrm{for}\ V=\gamma,\\
	 e\cot\theta_W & \textrm{for}\ V=Z^0.
	\end{cases}
\end{align}
In the SM at tree level, $\kappa_V=1$ and $\lambda_V=0$. We will focus only on the $C$- and $P$-conserving 
terms, discarding $\mathcal{L}_{WWV}^\textrm{nCP}$ in the following. In the photon case, 
the form factors are related to the static properties of the $W$ boson (namely the magnetic dipole 
$\mu_W$ and the electric quadrupole moment $Q_W$) through the relations~\cite{Hagiwara:1986vm} 
\begin{equation}
	\begin{aligned}
		\mu_W &= \frac{e}{2M_W}(1+\kappa_\gamma+\lambda_\gamma),\\
		Q_W &=- \frac{e}{M_W^2}(\kappa_\gamma-\lambda_\gamma).
	\end{aligned}	
\end{equation}
Following a notation analogous to the one used in Ref.~\cite{Argyres:1992vv} (see Fig.~\ref{fig:TGC} for the definition of the momenta), the $WWV$ vertex in momentum space can be written as
\begin{align}
	\Gamma_{\mu\alpha\beta}^V=-ig_{V} & \Big \{f(q^2) \left[2g_{\alpha\beta}p_\mu+4(g_{\alpha\mu}q_{\beta}-g_{\beta\mu}q_\alpha)\right]+2\Delta\kappa_V(q^2) (g_{\alpha\mu}q_{\beta}-g_{\beta\mu}q_\alpha)\notag\\
	&\left.+4\frac{\Delta Q(q^2)}{M_W^2}\left(p_\mu q_\alpha q_\beta-\frac{1}{2}q^2 g_{\alpha\beta}p_\mu\right)\right\},
\end{align}
with the $f(q^2)$ form factor connected to the renormalization of the charge, while $\Delta\kappa_V(q^2)$ and $\Delta Q_V(q^2)$, related to $\kappa_V$ and $\lambda_V$
through the expressions
\begin{equation}\label{eq:DefFFDelta}
	\begin{aligned}
		\Delta\kappa_V&=\kappa_V+\lambda_V-1 \equiv \Delta\kappa_V^{SM} + \Delta\kappa_V^{NP} \, ,\\
		\Delta Q_V&=-2\lambda_V \equiv \Delta Q_V^{SM}+\Delta Q_V^{NP} \, ,
	\end{aligned}	
\end{equation}
are designed to be zero at tree level in the SM. The SM one-loop
contributions can be found in
Refs.~\cite{Argyres:1992vv,Papavassiliou:1993ex,Fleischer:1991vq},
while the explicit calculation of $\Delta\kappa_V^{NP}$ and $\Delta
Q_V^{NP}$ in the case of vectorlike fermions is presented in
Appendix~\ref{app:VFtoTGC}. 

The quantity used by the experimental collaborations to show their results is the
deviation from the SM value of $\kappa_V$ at tree level, $\delta\kappa_V=\kappa_V-1$, 
which will correspond to a linear combination of $\Delta \kappa_V$ and $\Delta Q_V$, namely,
\begin{equation}\label{eq:deltakappa}
	\delta\kappa_V=\Delta\kappa_V+\frac{1}{2}\Delta Q_V,
\end{equation}
and this is the quantity we will be using throughout the paper.

\section{Models of colorless vectorlike fermions}
\label{sec:VLF}

For our study, we will consider two classes of colorless vectorlike
fermions: {\it (i)} a set of fermions in a unique SU(2)$_L$
representation, with no couplings to the Higgs boson allowed, and {\it
  (ii)} a set of at least two extra fermions in representations such
that a Yukawa term with the Higgs boson is allowed. In both cases, we
will assume that, due to some unspecified symmetry ${\cal G}$, all the
mixing between the vectorlike and the SM fermions is forbidden.


\subsection{Unmixed colorless vectorlike fermions}
\label{sec:MDegVF}

As already mentioned, we start adding to the SM particle content one vectorlike fermion $\Psi$, transforming under $SU(2)_L \times U(1)_Y$ as $\Psi \sim ({\bf 2j+1}, Y)$ and with mass $m_{\Psi}$. The Lagrangian is given by
\begin{align}
	\mathcal{L} = i\overline{\Psi}\gamma^\mu(\partial_\mu-ig W_\mu^aT^a-ig^\prime Y B_\mu)\Psi-m_{\Psi} \overline{\Psi}\Psi \, ,
\end{align}
where $T^a$ are the $2j+1$-dimensional generators of the $SU(2)_L$ Lie
algebra. An important consequence of considering a unique $SU(2)_L$
representation for all the $N_F$ vectorlike fermions is that the $\delta\kappa_V^{\Psi}$ form factor just depends on the
hypercharge and on the dimension $j$ of the $SU(2)_L$ representation and
not on the eigenvalues of the $T^3$ operator. This is shown explicitly
in Appendix~\ref{app:proof_Y_dependence}, from which we see that we can write
\begin{equation}\label{eq:eff_coupl}
	\delta\kappa_V^\Psi \propto F_j I(m_\Psi), ~~~~~F_j\equiv N_F\,Y\,\frac{2}{3} j (j+1) (2j+1),
\end{equation}
where $I(m_\Psi)$ is a loop factor that only depends on the vectorlike lepton mass $m_\Psi$. 
An equivalent statement is
that all the contributions to the $W^+W^-W^3$ TGC cancel out, leaving
only $W^+W^-B$ (with $B$ the hypercharge gauge boson). Integrating
numerically over the Feynman parameters of Eq.~(\ref{eq:VLF_TGC}) we
obtain $\Delta\kappa_V^\Psi$ and $\Delta\kappa_Z^\Psi$ as a function
of $\sqrt{s} = \sqrt{(2q)^2}$ (see
Appendix~\ref{app:VFtoTGC} for details).

In Fig.~\ref{fig:CVFIsoFCCee} we show the contour lines for $\delta\kappa_V^\Psi$ in the $(m_\Psi,\vert F_j\vert)$ plane for the 
four different center-of-mass energies $\sqrt{s}=m_H,\, 500\ \GeV,\, 1\ \TeV,$ and $3\ \TeV$. 
We observe that $\vert \delta \kappa_\gamma^\Psi \vert < \vert \delta \kappa_{Z^0}^\Psi\vert$ 
and they have opposite sign [see Eq.~(B.12) in 
Appendix~\ref{app:proof_Y_dependence}].
The typical values of $\vert \delta \kappa_V^\Psi \vert$ are smaller than a few $10^{-4}$.

\begin{figure}[t!]
\centering
\includegraphics[width=\textwidth]{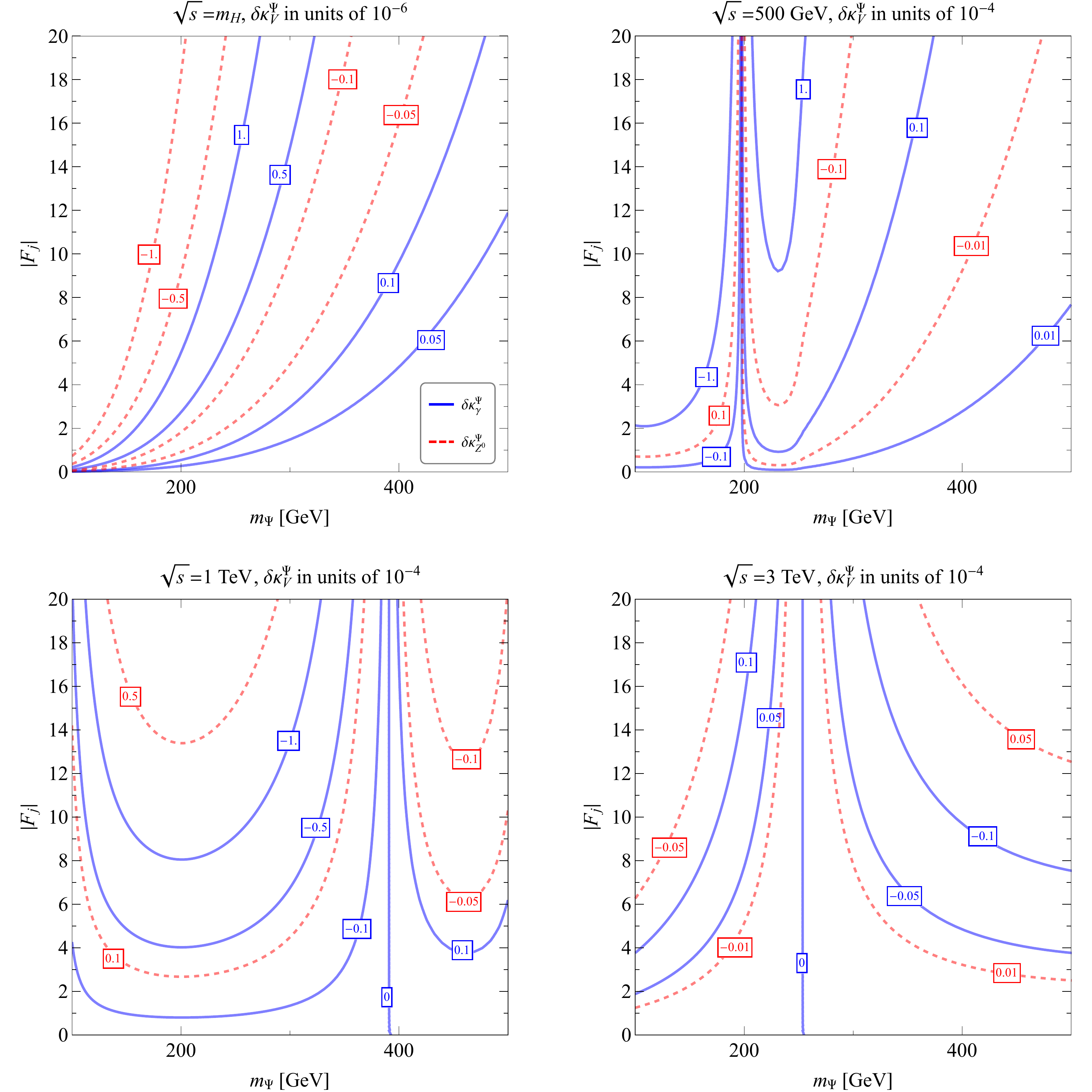}
\caption{Contour lines of $\delta\kappa_V^\Psi$ [see Eq.~(\ref{eq:deltakappa})] in the plane $(m_\Psi,\vert F_j\vert)$ for the models with unmixed
  vectorlike colorless fermions (vectorlike leptons)  at four different center-of-mass energies:
  $\sqrt{s}=m_H,\, 500\ \GeV,\,  1\ \TeV,$ and $3\ \TeV$.  For the definition of $F_j$, see Eq.~(\ref{eq:eff_coupl}). The full blue
  (dashed red) lines correspond to $V=\gamma$ ($Z^0$).}
\label{fig:CVFIsoFCCee}
\end{figure}

For fixed $\sqrt{s}$, the loop factor in Eq.~(\ref{eq:eff_coupl}) vanishes for $m_\Psi=m_{\Psi_1}$ and $m_\Psi=m_{\Psi_2}$, where $m_{\Psi_1,\Psi_2}$ are complicated functions of  $\sqrt{s}$. 
The general behavior of $\delta \kappa_\gamma^\Psi$ 
as a function of $m_\Psi$ is the following: it starts positive, vanishes for $m_\Psi=m_{\Psi_1}$, 
goes through a minimum (negative) value, increases again until it reaches zero for $m_\Psi=m_{\Psi_2}$, 
goes through a  maximum (positive) value, and then decreases again until it goes back to zero. Because of the flip in sign, $\delta \kappa_{Z^0}^\Psi$ has the opposite behavior. 
For $\sqrt{s}=m_H$, both cancellations occur for $m_\Psi<$ 100 GeV, so they do not appear in the plot. For 
$\sqrt{s}= 500$ GeV and $1$ TeV, we can only see in Fig.~\ref{fig:CVFIsoFCCee} 
the second cancellation at $m_{\Psi_2} \approx 200$
and 400 GeV, respectively, while for $\sqrt{s}=$ 3 TeV, we can see the first cancellation at 
$m_{\Psi_1} \approx 250$ GeV. Note that after the second cancellation the loop integral gets 
suppressed ($m_\Psi$ becomes too off shell for that specific center-of-mass energy) so to reach the 
same $\vert \delta \kappa_V^\Psi\vert$ one has to increase the effective coupling, 
{\it i.e.}, go to higher values of $\vert F_j \vert$.


\subsection{Mixed colorless vectorlike fermions}
\label{sec:MMixVF}

Let us now consider the case in which the colorless vectorlike fermions
transform in different $SU(2)_L \times U(1)_Y$ representations, such that an invariant 
Yukawa coupling with the Higgs boson is allowed. Since a general
discussion would be quite involved, we will consider two examples to
illustrate the impact of the future experiments measuring the
TGCs. Specifically, we will examine the two models studied
in Ref.~\cite{Almeida:2012bq}, corresponding to the addition of a singlet
and a doublet, and a doublet plus a triplet of fermions.


{\bf Doublet-singlet model}. We introduce a singlet Dirac fermion
$N = N_L + N_R$ with hypercharge $Y$ and a doublet Dirac fermion $L = L_L + L_R$
with hypercharge $Y-\frac{1}{2}$.\footnote{Notice that, although we
  use a notation suggesting heavier copies of a lepton doublet and
  right-handed neutrinos, we leave the hypercharge $Y$ of $N$
  unspecified. The case $Y=0$ corresponds, for example, to the situation
  studied in Ref.~\cite{Arvanitaki:2016xds}.} We will write explicitly the
components of the $L$ doublet as $L = (N_0, E)^T$ for the two
chiralities. The Lagrangian is given by
\begin{equation}\label{eq:L_doublet_singlet}
	\mathcal{L}_{2+1} = i \overline{L}  \slashed{D} L + i \overline{N}  \slashed{D} N - M_N \overline{N_R} N_L - M_L \overline{L_R} L_L - c \, \overline{N_R} H L_L - c' \, \overline{N_L} H L_R + h.c.
\end{equation}
With the hypercharge assignment we are considering, the electric charges of the various 
components are
\begin{equation}
	\begin{array}{ccl}
        E & \to & q_\chi  \equiv Y-1 \, ,\\
	N, N_0 & \to &  q_\omega \equiv Y
	\end{array}
\end{equation}
so that after electroweak symmetry breaking the Higgs introduces a
mixing between $N_0$ and $N$, while $E$ does not mix.

The three mass eigenstates $\omega_{1,2}$ and $\chi$  are defined as 
\begin{equation}
\omega = \pmatr{\omega_1 \\ \omega_2} = U_L^\dag \pmatr{N \\ N_0}_L + U_R^\dag \pmatr{N \\ N_0} _R, ~~~~ \chi = E_L + E_R\, ,
\end{equation}
with $U_{L/R}$ the unitary matrices that diagonalize the mass matrix
obtained from Eq.~(\ref{eq:L_doublet_singlet}) after electroweak
symmetry breaking. 

In terms of the mass eigenstates, the gauge Lagrangian can be written as
\begin{equation}
	\begin{aligned}
		\mathcal{L}_{\rm gauge}^{2+1}=& e\,q_\chi \bar{\chi}
                \gamma^\mu \chi A_\mu + e q_\omega \,\overline{\omega}
                \gamma^\mu \omega A_\mu
                -\frac{1}{2}\Big((2Y-1)g^\prime s_W+g c_W
                \Big)\overline{\chi}\gamma^\mu\chi Z_\mu
                \\ &+\overline{\omega}\left[ U_L^\dagger
			\begin{pmatrix}
				-Y\,g^\prime s_W & 0 \\
				0 & \frac{1}{2}\big(g \, c_W-(2Y-1)g^\prime s_W \big)
			\end{pmatrix}
			U_L P_L+(L \to R)
		\right]\gamma^\mu \omega Z_\mu,\\
		& + \frac{g}{\sqrt{2}}\overline{\omega}\gamma^\mu [U_L^\dagger P_L+U_R^\dagger P_R](0\quad W_\mu^+)^T \chi \, ,
	\end{aligned}
\end{equation}
where $g$ and $g^\prime$ are the usual SM gauge couplings, $s_W=\sin \theta_W$ and $c_W=\cos \theta_W$.

\begin{figure}[t!]
	\centering
	\includegraphics[width=\textwidth]{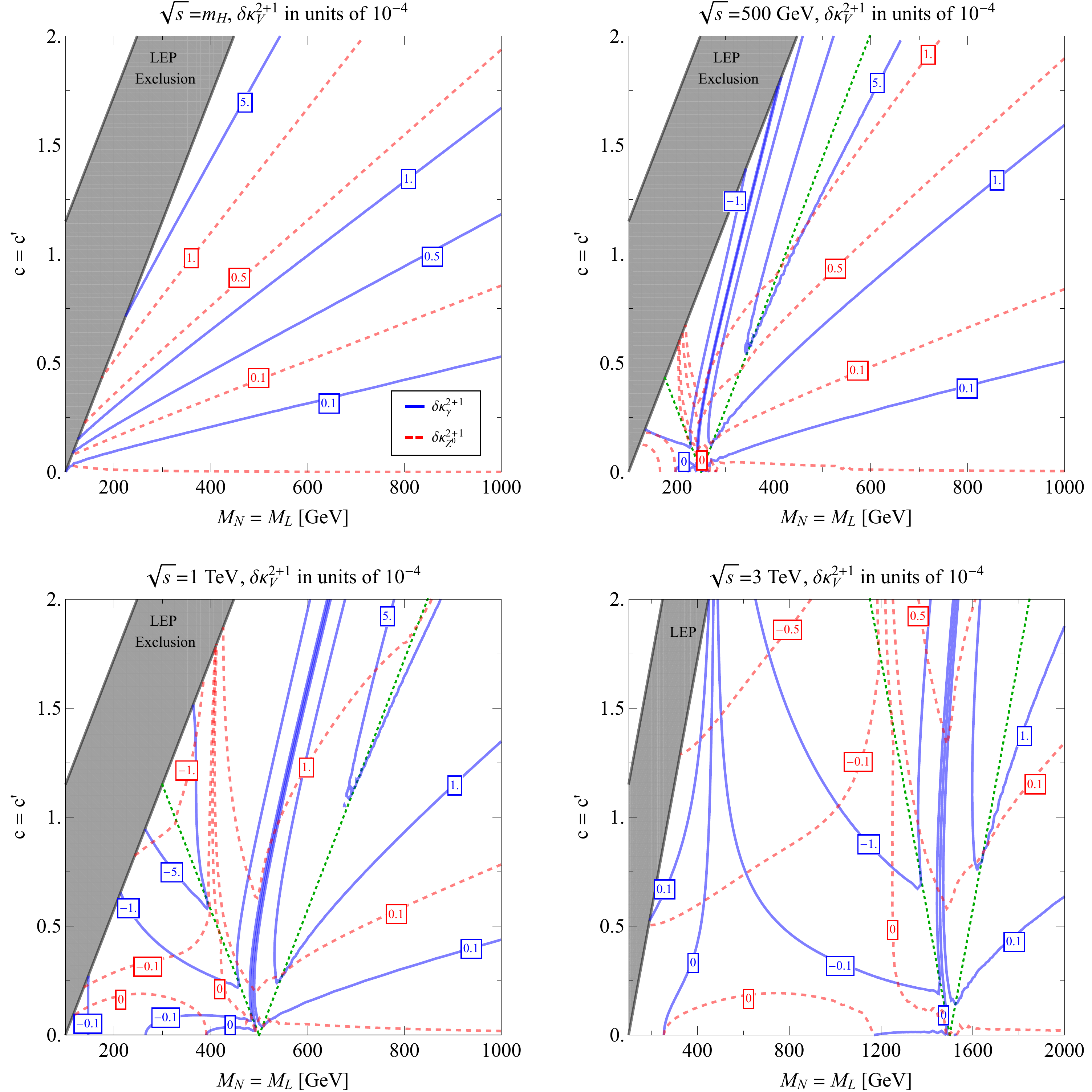}
	\caption{Isocontour lines of the deviations $\delta\kappa_V^{2+1}$ from the SM couplings 
		in the plane $(M_N=M_L,c=c^\prime)$  for the 
		vectorlike colorless fermion doublet-singlet model  at 
		four different center-of-mass energies: $\sqrt{s}=m_H,\, 500\ \GeV,\,  1\ \TeV,$ and $3\ \TeV$.
		We have chosen $Y=1$, so $\omega_1$ and $\omega_2$ are charged, whereas $\chi$ is neutral.
		The full blue (dashed red) lines correspond to $V=\gamma$ ($Z^0$).
		The dotted green lines correspond to the physical masses $m_{\omega_1}$ 
		and $m_{\omega_2}$, for $M_N=M_L=\sqrt{s}/2$.
		}
	\label{fig:D+SIsoFCCee}
\end{figure}
\noindent Having established our model, we proceed to compute the
one-loop contributions of the new vectorlike fermions to the 
TGCs. Using the general result for the one-loop contribution, given in Appendix
\ref{app:VFtoTGC}, we compute the $\Delta\kappa_V^{2+1}$
and $\Delta Q_V^{2+1}$ form factors for this model. 
Note that the $W^+W^-Z^0$ vertex gets an additional correction 
with respect to the $W^+W^-\gamma$ one, due to the mixing between the doublet and 
the singlet.

In Fig.~\ref{fig:D+SIsoFCCee}, we show the contour lines for $\delta \kappa_V^{2+1}$ in the $(M, c)$ plane, 
where $M = M_L = M_N$ and $c' = c$, for the same four center-of-mass energies as before. Assuming $c$ real, 
the mass spectrum is $m_\chi=M$, $m_{\omega_1, \omega_2}=\vert M \pm 2c \, v\vert$, where $v=175$ GeV is 
the SM Higgs vacuum expectation value. For an illustration, we have chosen the case $Y=1$, so $\omega_1$ and $\omega_2$ are 
particles with charge 1 that participate in both $\delta \kappa_{\gamma}^{2+1}$
and $\delta \kappa_{Z^0}^{2+1}$, whereas $\chi$ is a neutral fermion and so it only contributes 
to the latter. For a fixed coupling $c=c'$, $\delta \kappa_\gamma^{2+1}$ has the following behavior as a function of 
$M=M_L = M_N$. It starts positive when, for a given center-of-mass energy, all vectorlike fermion masses are irrelevant 
for the loop function. Then, it decreases as the lowest fermion mass starts to play a role, until it reaches
a minimum at $m_{\omega_1}=|\sqrt{s}/2-2cv|$; next, it increases when the next massive vectorlike fermion 
starts to contribute and passes again through zero before reaching a maximum at 
$m_{\omega_2}=\sqrt{s}/2+2cv$. As $M_N$ continues to increase, $\delta \kappa_\gamma^{2+1} \to 0$
as we approach the decoupling limit.  The behavior of $\delta \kappa_{Z^0}^{2+1}$ is somewhat similar 
but a bit more involved at lower values of $M_N$ due to the mixing between $\omega_{1,2}$. Also, 
as $M_N$ increases, the contribution of the neutral vectorlike fermion, $\chi$, appears,  giving  
rise to the maximum value for $\delta \kappa_{Z^0}^{2+1}$ at $M_N=m_\chi$.
Here, again, the typical values of $\vert \delta \kappa_V^{2+1} \vert$ are smaller than a few $10^{-4}$.
The green dotted lines that can be seen 
on the $\sqrt{s}= 500$ GeV and 1 TeV panels correspond to the values of $m_{\omega_1}$ and $m_{\omega_2}$ computed 
with $M=\sqrt{s}/2$. At the other center-of-mass energies, these masses lie outside of the plot range.
 

{\bf Triplet-doublet model}. We will now add to the SM particle content a Dirac SU(2)$_L$ doublet $L = L_L + L_R$ and
a Dirac triplet $T = T_L + T_R$, with hypercharges $Y$ and $Y-\frac{1}{2}$, respectively. The total Lagrangian is given by
\begin{align}
	\mathcal{L}_{3+2}=& \, i \overline{L} \slashed{D} L + i \overline{T} \slashed{D} T - M_L \overline{L_L} L_R - M_T \overline{T_L} T_R - c \, \overline{L_L} T_R H - c' \, \overline{L_R} T_L H + h.c. \, ,
\end{align}
where the doublet and triplet fermions are written as
\begin{equation}
L = 
\begin{pmatrix}
	N_0 \\
	E
\end{pmatrix}\, , ~~~~ T = 
\begin{pmatrix}
	\frac{T_a}{\sqrt{2}} & T_b \\
	T_c & -\frac{T_a}{\sqrt{2}}
\end{pmatrix} \, .
\end{equation}
With the hypercharge assignment we are considering, the electric charges of the various components read
\begin{equation}
	\begin{array}{crl}
	T_c & \to & q_\chi \equiv Y - \frac{3}{2}   \, , \\
	T_a, E & \to & q_\xi \equiv Y - \frac{1}{2}   \, ,\\
	T_b, N_0 & \to & q_\omega \equiv Y + \frac{1}{2}  \, ,
	\end{array}
\end{equation}
in such a way that, after electroweak symmetry breaking, there is a mixing between $T_a$ and $E$, as well as between $T_b$ and $N_0$. Defining the mass eigenstates as
\begin{gather}
\omega = \pmatr{\omega_1 \\ \omega_2} = U_L^\dag \pmatr{N_0 \\ T_b}_L + U_R^\dag \pmatr{N_0 \\ T_b}_R,  ~~~ \xi = \pmatr{\xi_1 \\ \xi_2} = V_L^\dag \pmatr{E \\ T_a}_L + V_R^\dag \pmatr{E \\ T_a}_R  , \nonumber\\ 
\vspace{0.3cm} \\
\chi = T_{cL} + T_{cR}, \nonumber
\end{gather}
the gauge Lagrangian can be written as
\begin{align}
	\mathcal{L}_{3+2}=& e \, q_\chi \, \overline{\chi}\gamma^\mu\chi A_\mu + e \, q_\omega \,\overline{\omega}\gamma^\mu \omega A_\mu +  e \, q_\xi \,\overline{\xi}\gamma^\mu \xi A_\mu -  \left( q_\xi \, g^\prime \, s_W + g \, c_W \right) \overline{\chi}\gamma^\mu\chi Z_\mu \notag \\
	+ & \bar{\omega}\left[
			U_L^\dagger
			\begin{pmatrix}
				\frac{g}{2}c_W - Y \, g^\prime s_W & 0 \\
				0 & g c_W - q_\xi \, g^\prime s_W
			\end{pmatrix}
			U_L P_L+(L \to R)
		\right]\gamma^\mu \omega Z_\mu\notag\\
		+ & \overline{\xi}\left[
			V_L^\dagger
			\begin{pmatrix}
				-\frac{g}{2}c_W - Y\,g^\prime s_W & 0 \\
				0 & -q_\xi g^\prime s_W
			\end{pmatrix}
			V_L P_L+(L \to R)
		\right]\gamma^\mu \xi Z_\mu  \notag \\
		+ & g\,(\overline{\omega} \,\,\,  \overline{\xi} \,\,\,  \overline{\chi})\gamma^\mu \left[
					\begin{pmatrix}
						0_{2\times 2} & W_\mu^+ \, U_L^\dagger \, V_L^\prime & 0_{2\times 1} \\
						W_\mu^- \, V_L^{\prime\dagger} \, U_L & 0_{2\times 2} & V_L^\dagger \, \widetilde{W}_\mu^{+T}\\
						0_{1\times 2} & \widetilde{W}_\mu^{-T} \,  V_L & 0
					\end{pmatrix}
					P_L + (L\to R)				
			\right]
			\begin{pmatrix}
				\omega\\
				\xi\\
				\chi
			\end{pmatrix},
\end{align}
where $\widetilde{W}_\mu^\pm = (0 \,\,\, W_\mu^\pm)$ and
\begin{align*}
	V_L^\prime =
	\frac{1}{\sqrt{2}}
	\begin{pmatrix}
		V_{L11} & V_{L12}\\
		\sqrt{2} V_{L21} & \sqrt{2} V_{L22}
	\end{pmatrix}.
\end{align*}

\begin{figure}[t!]
\centering
\includegraphics[width=\textwidth]{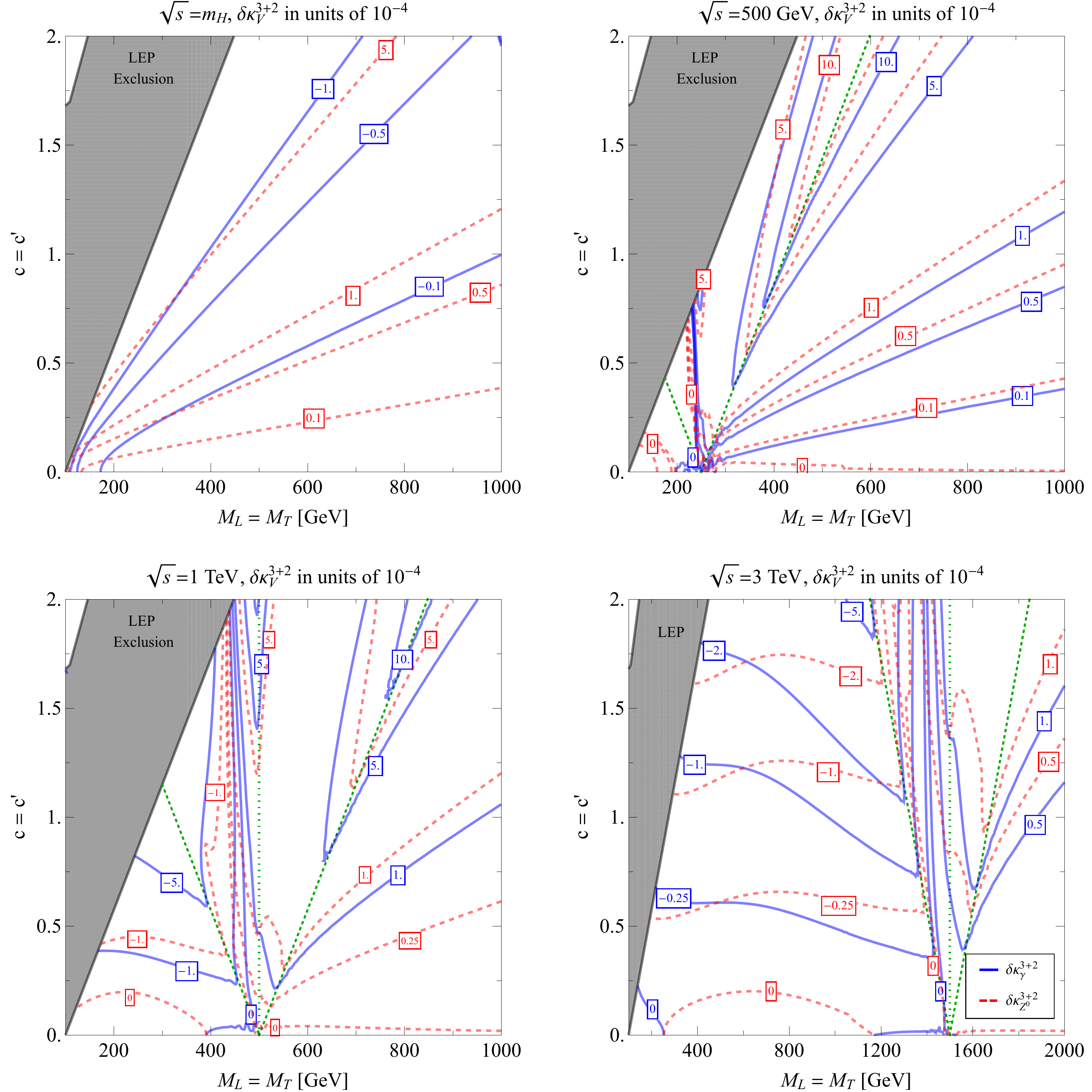}
\caption{Isocontour lines of the deviations $\delta\kappa_V^{3+2}$ from the SM couplings 
in the plane $(M_L=M_T,c=c^\prime)$  for the 
vectorlike colorless fermion triplet-doublet model  at 
four different center-of-mass energies: $\sqrt{s}=m_H,\, 500\ \GeV,\,  1\ \TeV,$ and $3\ \TeV$.
We have chosen $Y=1/2$, so there are three charged states and two neutral ones.
The full blue (dashed red) lines correspond to $V=\gamma$ ($Z^0$), and 
the dotted green lines correspond to the physical masses $m_{\omega_1}$, $m_{\omega_2}$,
and $m_{\chi}$.}
\label{fig:T+DIsoFCCee}
\end{figure}

In Fig.~\ref{fig:T+DIsoFCCee}, we show the isocontour lines for the 
$\delta \kappa_V^{3+2}$ combinations for this model in the plane 
$M_L=M_T$ vs $c=c'$ for the same four different center-of-mass energies 
as before. In this case the physical mass spectrum is $m_\chi=\sqrt{s}/2$,
$m_{\omega_1}=\vert \sqrt{s}/2- 2c \, v\vert$, $m_{\omega_2}=\sqrt{s}/2+ 2c \, v$, 
$m_{\xi_1}=\vert \sqrt{s}/2- \sqrt{2}c \, v\vert,$ and $m_{\xi_2}=\sqrt{s}/2+ \sqrt{2}c \, v$. The green dotted lines 
that can be seen on the $\sqrt{s}= 500$ GeV and 1 TeV panels correspond to the values of the charged particle masses 
$m_{\omega_1}$, $m_{\omega_2}$, and $m_\chi$. 
At the other center-of-mass energies these masses lie outside of the plot range.

Here, we show the case $Y=1/2$, so $\chi$, $\omega_1$, and $\omega_2$ are 
charged particles that participate of both $\delta \kappa_{\gamma}^{3+2}$
and $\delta \kappa_{Z^0}^{3+2}$, whereas $\xi_1$ and $\xi_2$ are neutral fermions and  only contribute 
to the latter. Here, the typical values of $\vert \delta \kappa_V^{3+2} \vert$ can get about an 
order of magnitude larger than in the previous models, but the form factors are always smaller than a few $10^{-3}$.

For a fixed coupling $c$, $\delta \kappa_\gamma^{3+2}$  as a function of 
$M_L$ has the same general behavior as for the doublet-singlet model. 
It goes through a minimum at $m_{\omega_1}$, and through a maximum 
at $m_{\chi}$ and $m_{\omega_2}$. This can be best seen on the panel for 
$\sqrt{s}=$ 1 TeV. 
The behavior of $\delta \kappa_{Z^0}^{3+2}$ is somewhat similar 
but even more involved than the previous mixed case because now 
we have five particles coupling to the $Z^0$ so in addition to 
the charged particle peaks, we also have peaks for the neutral 
particles. We note that in this case $\vert\delta \kappa_{Z^0}^{3+2}\vert \sim 
\vert\delta \kappa_{\gamma}^{3+2}\vert$ and sometimes even a bit larger.

\begin{figure}[t!]
\centering
\center{Unmixed colorless vectorlike  fermion scenario}
\vspace{0.1cm}
\includegraphics[width=0.95\textwidth]{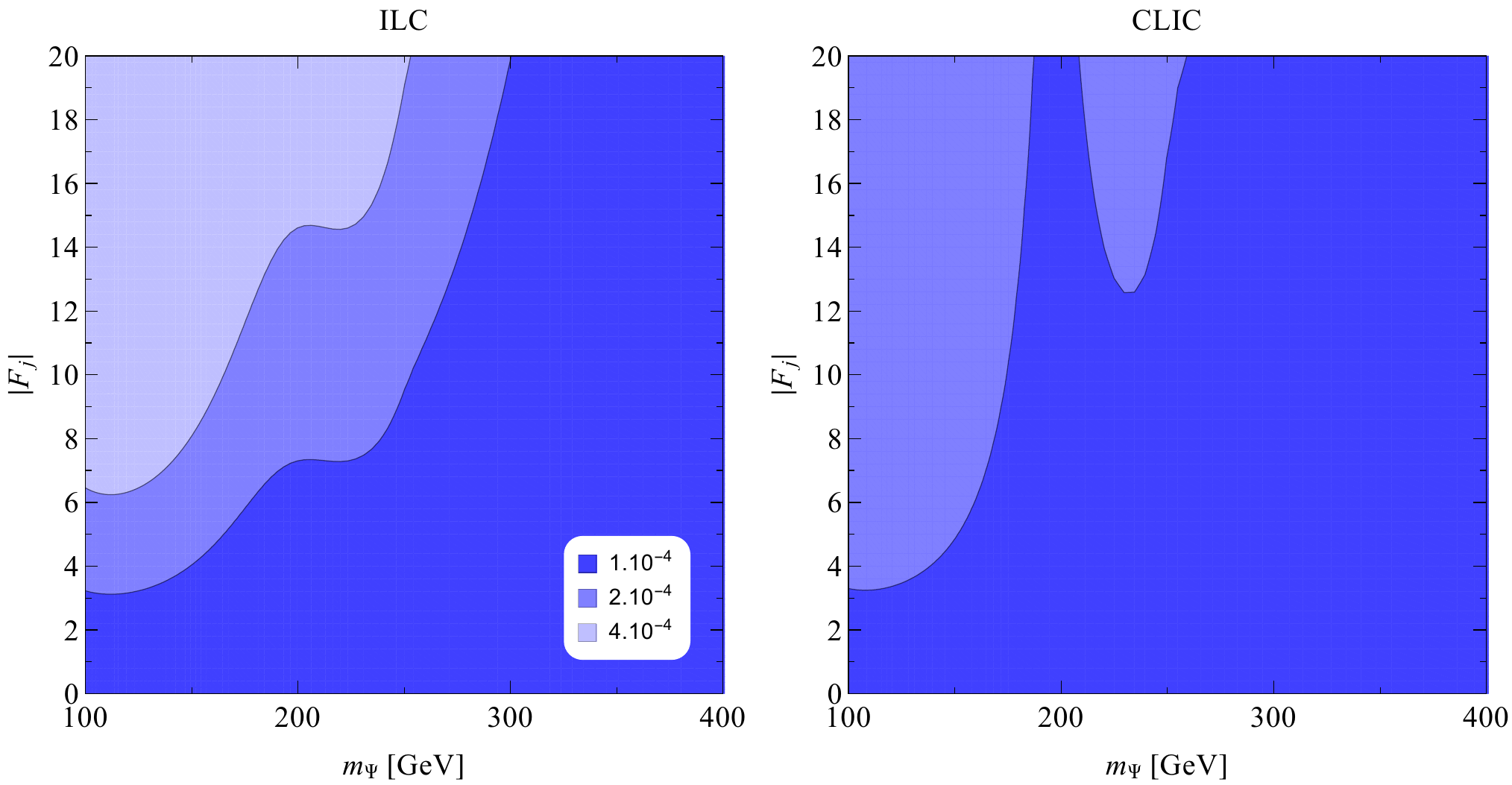}
\caption{Possible TGC reach to probe the parameters of the unmixed vectorlike colorless fermion models
	by combining different center-of-mass energies at the ILC ($\sqrt{s}=500,800,1000$~GeV)
	and the CLIC ($\sqrt{s}=500,1400, 3000$~GeV) facilities. 
	We assume the same three different sensitivities for 
	$\delta \kappa_\gamma$ and $\delta \kappa_{Z^0}$
	at all center-of-mass energies considered: $4\times 10^{-4}$, $2\times 10^{-4}$  and $1\times 10^{-4}$. 
	The regions of accessibility were computed at 95.45\% C.L.. See the text for more details.}
\label{fig:LimCVF}
\end{figure}

\section{TGC constraints on vectorlike colorless fermion models}
\label{sec:Constraints}
%

We move now to estimate the possible future constraints that can be imposed on
vectorlike colorless fermion models by TGC measurements at future $e^+e^-$ accelerator facilities such as
the proposed FCC-ee \cite{dEnterria:2016sca}, ILC \cite{Baer:2013cma}, and 
the CLIC \cite{Dannheim:2012rn}.
For the FCC-ee experiment we considered the following center-of-mass energies:  
$\sqrt{s}=m_Z,\ m_H,\ 2m_Z$ and $2m_t$~\protect\cite{dEnterria:2016sca}, 
for the ILC: $\sqrt{s}=500,\ 800,$  and 1000~GeV~\protect\cite{Baer:2013cma}; and 
for the CLIC (in the so-called scenario A): 
$\sqrt{s}=500,\ 1400,$ and 3000~GeV~\protect\cite{Dannheim:2012rn}.

\sloppy We do this for each of the models addressed in this paper 
by minimizing a combined $\chi^2(\delta \kappa_Z,\delta \kappa_\gamma;\sqrt{s_i}$) assuming the following three
different benchmark sensitivities 
for both TGCs:
$4\times 10^{-4}$, $2\times 10^{-4}$, and
$1\times 10^{-4}$~\cite{Ellis:2015sca,Barklow:2015tja}.
We assume the same benchmarks for all facilities at all center-of-mass energies.

In Fig.~\ref{fig:LimCVF}, we show the regions on the plane ($m_\Psi$,
$\vert F_j \vert$) of the unmixed vectorlike model that can be probed
at 2$\sigma$ C.L. by combining the various center-of-mass energies at these
accelerators.  Because of the relatively low center-of-mass energies proposed for the
FCC-ee, it can only probe a very limited range of $m_\Psi \lesssim 200$ GeV for 
$\vert F_j\vert \gtrsim (1-4)$ at 2$\sigma$ C.L. if 
the sensitivity is at least $1\times 10^{-4}$.  This is why we do not show this case 
on Fig.~\ref{fig:LimCVF}. 
The ILC  will be able to test  $m_\Psi \lesssim 250$ GeV ($m_\Psi \lesssim 300$ GeV)
for $\vert F_j\vert \gtrsim 16$
if a sensitivity of $2\times 10^{-4}$ ($1\times 10^{-4}$) can be achieved.
At the CLIC, the reach is somewhat reduced, 
as, for instance, no region is accessible at 2$\sigma$ C.L. 
even for a sensitivity of $2\times 10^{-4}$ for $|F_j|<20$.
Note that the CLIC is less sensitive to the unmixed colorless vectorlike scenario than the ILC due to its higher 
center-of-mass energies as explained by the following reasoning. As can be seen in 
Fig.~\ref{fig:CVFIsoFCCee}, the contribution to TGCs is higher 
when $\sqrt{s}$ is close to the vectorlike fermions mass threshold, but the heavier the fermions are, the 
smaller the TGC deviation is in general. 
Deviations at the $\mathcal{O}(10^{-4})$ level are typically caused by 
particles below the TeV scale, and thus having a lower center-of-mass energy leads to better sensitivity.

\begin{figure}[t]
\centering
\includegraphics[width=0.95\textwidth]{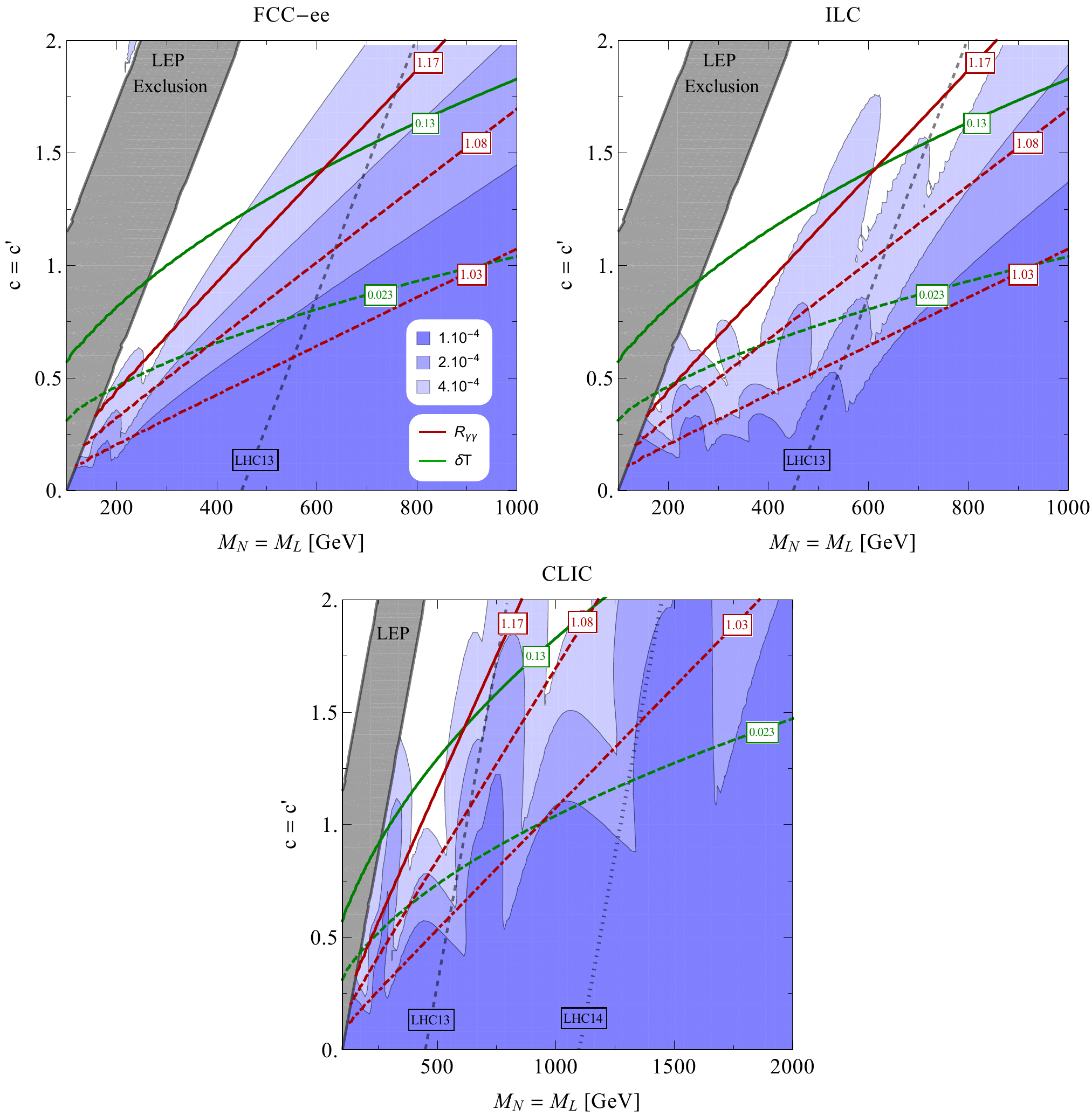}
\caption{Possible TGC reach to probe the parameters of the doublet-singlet vectorlike colorless
fermion model with $Y=1$, by combining different center-of-mass energies at the FCC-ee, the ILC and 
the CLIC facilities at 2$\sigma$ C.L.. We also show the current 
limits from $H\to \gamma \gamma$ ($R_{\gamma \gamma}$, full red line) and
electroweak precision measurements ($\delta T$, full dark green line) as well as the 
possible future sensitivities of
$R_{\gamma \gamma}$ assuming an uncertainty of 
8\% (dashed red line) or 3\% (dotted-dashed red line) 
and of $\delta T$ (dashed dark green line). The gray region has been 
excluded by LEP~\cite{Achard:2001qw}, while the black dashed (dotted) lines 
correspond to the LHC current limit (future sensitivity).}
\label{fig:LimDpS}
\end{figure}

In Fig.~\ref{fig:LimDpS}, we show the regions on the plane ($M_N=M_L$, $c=c'$) of the doublet-singlet 
model with $Y=1$ that can be explored by the FCC-ee, ILC, and CLIC at 2$\sigma$ C.L.. This is performed 
as before, that is, by combining the $\chi^2$ at the center-of-mass energies of each facility. 
For comparison, we also show the current limits one can obtain from $H\to \gamma \gamma$ 
($R_{\gamma \gamma}$, full red line; see, e.g., Ref.~\cite{Carena:2012xa}) and
electroweak precision measurements ($\delta T$, full dark green line) as well as the 
effect of  a future possible improvement on the uncertainty on 
$R_{\gamma \gamma}$ to 8\% (dashed red line) or 3\% (dotted-dashed red line) 
and on the uncertainty on $\delta T$ (dashed dark green line).
These future prospects on the uncertainties were taken from Refs. \cite{Baer:2013cma,Baak:2014ora};
for comparison we show the same $\delta T$ and $R_{\gamma\gamma}$ sensitivities for all proposed facilities.
The region in gray was excluded by LEP searches for neutral and charged leptons~\cite{Achard:2001qw}.

At present, $R_{\gamma \gamma}$  excludes more of the parameter space of the 
doublet-singlet model than $\delta T$ if $M_N \lesssim 600$ GeV,
but for larger values of $M_N$, $\delta T$ is more restrictive.
We see that at the FCC-ee one can have the sensitivity to probe and exclude 
a larger region of the parameter space, which can only be comparable to a future sensitivity 
on $R_{\gamma \gamma}$  of 8\% or better, if one can reach a sensitivity of 
$\sim 1.5\times 10^{-4}$ on the TGCs. 
Here since the center-of-mass energies that we 
have combined are comparatively low, the peak structure only appears around 
$M_N \sim 180$ GeV, the rest of the exclusion region being quite smooth.
At the ILC, because the center-of-mass energies are higher, the exclusion region 
is more complicated due to the maxima and minima that appear for the different masses 
of the vectorlike fermions that run in the loop functions at different $\sqrt{s}$. 
In general, the ILC can exclude the same regions probed by the FCC-ee but, 
most of the parameter space, requiring a  less challenging sensitivity to the TGCs.

\sloppy Although the CLIC involves even higher center-of-mass energies, it loses some 
sensitivity for $M_N\sim 700$ GeV because of the peaks structure. Nevertheless, it can 
test the regions $800\lesssim M_N/{\rm GeV}\lesssim 1400$ and $1600\lesssim M_N/{\rm GeV}\lesssim 1900$ 
for a TGC sensitivity of $1\times 10^{-4}$. Such region could only be inspected by 
a $R_{\gamma\gamma}$ or a $\delta T$ measurement with 2\%-3\% uncertainty.

\begin{figure}[t!]
\centering
\includegraphics[width=0.95\textwidth]{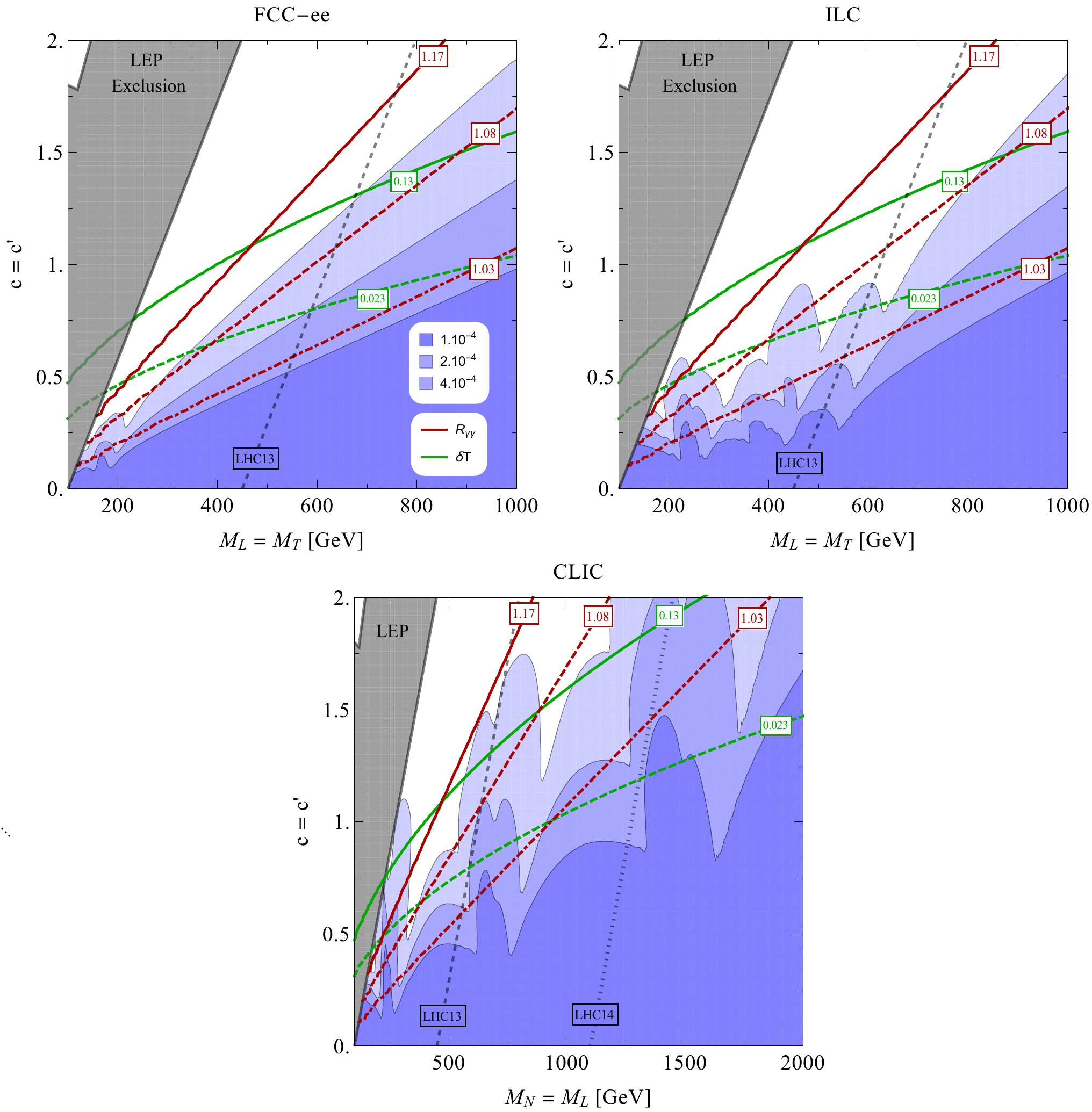}
\caption{Same as Fig.~\ref{fig:LimDpS} but for the triplet-doublet vectorlike colorless 
fermion model with $Y=1/2$.}
\label{fig:LimTpD}
\end{figure}

Finally, in Fig.~\ref{fig:LimTpD} we show the regions on the plane
($M_L=M_T$, $c=c'$) of the triplet-doublet model with $Y=1/2$ that can be explored at
2$\sigma$ C.L. by the FCC-ee, ILC, and CLIC, again combing the same 
center-of-mass energies as before. In this case, the FCC-ee can explore a region than can only 
be attainable by measuring  $R_{\gamma \gamma}$ with an uncertainty of at least 3\%
if the TGC sensitivity is $2\times 10^{-4}$, while the ILC is a bit better except  
for $M_L \lesssim 250$ GeV. As before CLIC is, in general, less sensitive for $M_L\lesssim 700$ GeV
because of the peak structure but becomes more sensitive for higher masses,
probing the model down to regions where even a very aggressive measurement of $R_{\gamma \gamma}$ would not reach.  

Let us conclude with some remarks about the limits from direct searches at the LHC. As shown, 
for instance, in Refs.~\cite{Arvanitaki:2016xds,Beauchesne:2017ukw}, the collider signatures 
of the doublet-singlet model are very similar to those of electroweakinos in minimal supersymmetry 
models. Moreover, we expect the limits for the other representations not to be too different. 
Current lower bounds can be found in Ref.~\cite{Sirunyan:2017zss} and are of order $150$ GeV for 
the lightest neutral state and of order $450$ GeV for the heavier states. Future sensitivities 
have been estimated in Ref.~\cite{ATLAS:fut}; with a luminosity of $3000$ fb$^{-1}$ 
(at $\sqrt{s} = 14$ TeV), the lower bound on the lightest neutral mass becomes $400$ GeV, 
while the lower bound on the heavier states becomes $1.1$ TeV. We included the current limit 
(dashed black line) and future sensitivity (dotted black line) in Figs. \ref{fig:LimDpS} and 
\ref{fig:LimTpD}. As can be seen, even considering the future LHC reach, there are regions not 
probed by the LHC that will be probed by TGCs searches.

\section{Conclusions}\label{sec:conc}

We have studied vectorlike colorless fermions contributions to the
triple gauge couplings $W^+ W^- \gamma$ and $W^+ W^- Z^0$ in the
context of two classes of models.  First, we considered the unmixed case, in which an arbitrary 
set of fermions in a given representation of SU(2)$_L$ cannot couple to the SM Higgs boson.
Second, we considered the mixed case, where the vectorlike fermion fields transform as different 
representations of SU(2)$_L$  allowing for invariant Yukawa couplings with the Higgs boson. 
In the latter case, we studied two concrete situations: the doublet-singlet model, 
where three new vectorlike physical particles are introduced,
and the triplet-doublet model, where five new vectorlike physical particles 
appear.

We established that the contributions of the above vectorlike fermion models 
to the combination of the form factors,
$\delta \kappa_V, \, V=\gamma,Z^0$, used by the experimental collaborations, 
have several minima and maxima as a function of the mass parameters of the model. 
Since to go from a negative minimum to a positive maximum one has to cross zero, this also implies that 
there are values of the mass parameter for which $\delta \kappa_V \to 0$.
These maxima and minima  will depend on the center-of-mass energy considered, and how close one is to a
physical particle which contributes to the TGC loop function being on the mass shell.

In the case of the unmixed vectorlike colorless fermion model, we have assumed that all fermions, 
independent of how many multiplets of a given representation,  are degenerate in mass ($m_\Psi$).  
Since $\vert\delta \kappa_\gamma\vert$ starts large when $m_\Psi\ll \sqrt{s}/2$, and we expect a
maximum at $m_\Psi \sim \sqrt{s}/2$, there are, in general, two  values of $m_\Psi$, for a given  
$\sqrt{s}$, where $\delta \kappa_V \to 0$.

For the doublet-singlet and triplet-doublet model the minima and maxima for $\delta \kappa_\gamma$ 
($\delta \kappa_{Z^0}$) as a function of $M_L$, the mass parameter, correspond to the values of the 
charged (all) physical particles of the model, which clearly depend on $\sqrt{s}$ and the hypercharge 
$Y$, which defines the charges of the particles. 

We made an assessment of the sensitivity of the proposed future precision test accelerators, the 
FCC-ee, ILC, and CLIC to the parameters of these models assuming they will be able to constrain
$\delta \kappa_V\sim\mathcal{O}(10^{-4})$ at different $\sqrt{s}$. 
Using the same benchmark sensitivities
for all accelerators allowed us to clearly see the effect of the different center-of-mass energy combinations.
For the FCC-ee experiment we considered the following center-of-mass energies:  
$\sqrt{s}=m_Z,\ m_H,\ 2m_Z$ and $2m_t$. For the ILC: $\sqrt{s}=500,\ 800,$ and 1000~GeV,
and for the CLIC (in the so-called scenario A): $\sqrt{s}=500,\ 1400,$ and 3000~GeV.

Only for the unmixed vectorlike colorless fermion case, the FCC-ee is
definitely not as capable to probe the model as the ILC or the
CLIC. However, for both mixed vectorlike models we have examined, the
ILC is generally better than the FCC-ee but not as powerful as the CLIC
at larger values of the mass parameters $M_N$ or $M_L$. This is
because the $\sqrt{s}$ used by FCC-ee are all quite low, making the
exclusion region basically insensitive to the maxima and minima caused
by the physical particle masses. 
For the ILC, the gaps between the center-of-mass energies and their high values 
exhibit some synergy that helps to improve the sensitivity in a large region of the parameter.  
This also happens for the CLIC, but since the center-of-mass energies are more spread out,
there is an overall decrease in sensitivity to the model parameters 
for $M_N$, $M_L \lesssim$ 700 GeV, with respect to the ILC. However, for higher masses (due 
to the 3000 GeV center-of-mass energy contribution) we have again an increase of sensitivity 
because heavier vectorlike fermion physical masses come into play.

It is also important to note that if one is able to achieve  ${\cal O}(10^{-4})$ sensitivity on TGCs with 
the FCC-ee, ILC, or  CLIC one will be able to use them to do precision measurements that surpass the 
sensitivities of the oblique parameters or $H \to \gamma \gamma$ even assuming a considerable improvement 
of the latter measurements in these new machines.

\begin{acknowledgments}
P. M. thanks the Universidade de S\~{a}o Paulo for the kindliest hospitality. This work was supported by 
Funda\c{c}\~ao de Amparo \`a Pesquisa do Estado de S\~ao Paulo (FAPESP) and Conselho Nacional de Ci\^encia 
e Tecnologia (CNPq). This project has also received partial funding from the European Union's Horizon 2020 
research and innovation program under the Marie Sklodowska-Curie Grants No. 674896 and No. 690575. 
Fermilab is operated by Fermi Research Alliance, LLC, under Contract No. DE-AC02-07CH11359 with the US Department of Energy.
\end{acknowledgments}

\appendix

\section{Vectorlike fermion vontribution to triple gauge couplings}\label{app:VFtoTGC}

\begin{figure}[ht]
  \centering
	\includegraphics[scale=0.45]{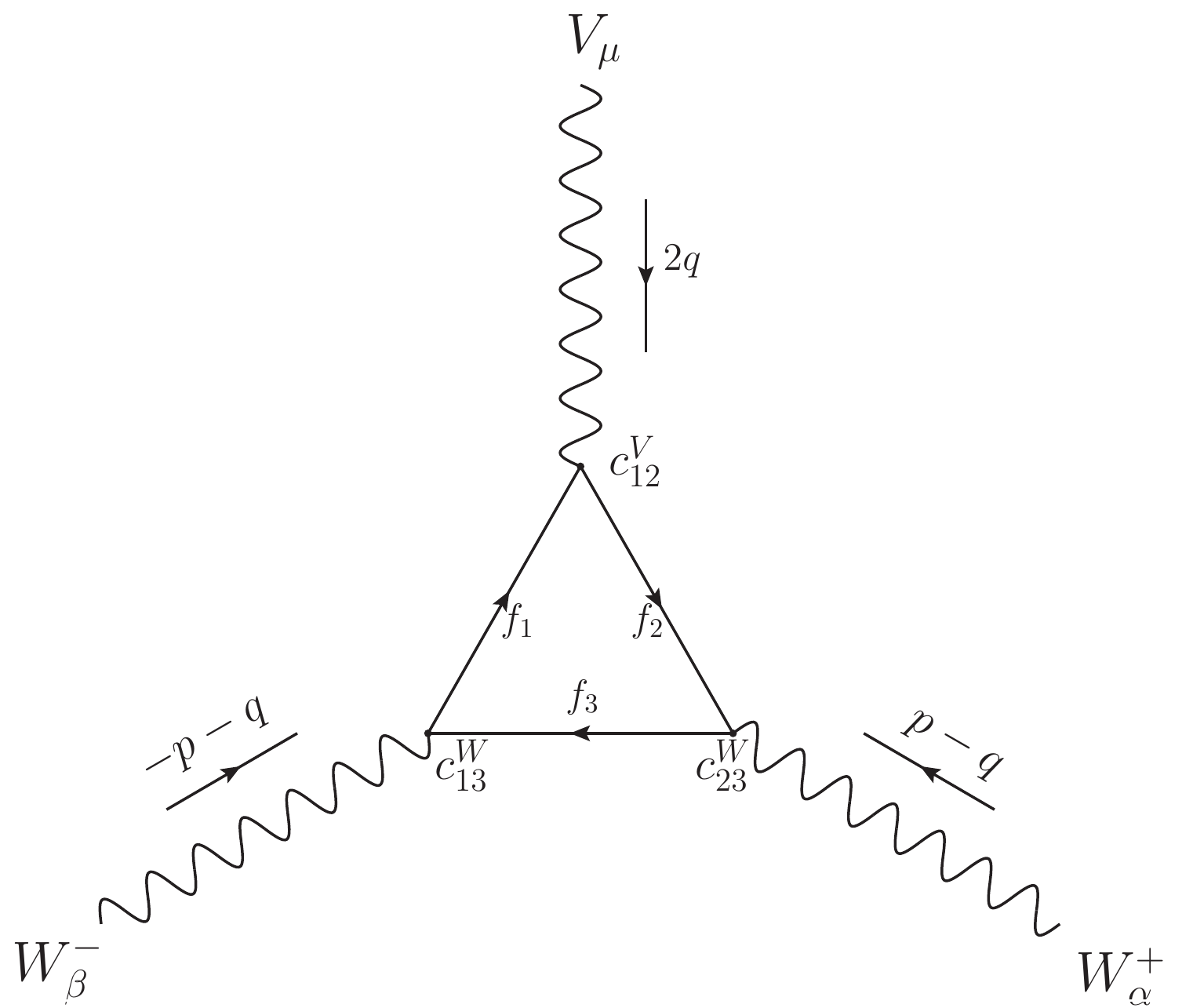}
	\caption{Vectorlike fermions contribution to TGC.}
	\label{fig:TGCVF}
\end{figure}

The one-loop correction to the TGCs coming from a set of $N_F$ vectorlike fermions
can be obtained from the diagram in  Fig.~\ref{fig:TGCVF}. Here, we will keep as general as possible, by supposing
that three different fermions run into the loop, $f_i$, $i=\{1,2,3\}$, with masses $m_i$ and generic couplings between them and the gauge bosons, $c_{ij}^B$, where $i,j=\{1,2,3\}$ and $B=\{\gamma,W,Z\}$. Proceeding in a standard way, we find the $\Delta\kappa_V^{NP}$ and
$\Delta Q_V^{NP}$ form factors,
\begin{subequations}\label{eq:VLF_TGC}
	\begin{align}
		\Delta\kappa_V^{NP}&=-N_F\frac{c_{12}^V c_{23}^W c_{31}^W}{8\pi^2g_V}\int_0^1dx\int_0^1dy\ \frac{x}{\widetilde{\Lambda}}\left\{\frac{4q^2}{M_W^2}x^2(3x-2)y(1-y)+x^2(x-1)\right.\notag\\
		&\qquad\qquad\qquad\qquad\left.+(R_1-R_2)xy(x-1)+(R_3-R_1)x(x-1)+\sqrt{R_1R_2}x\right.\notag\\
		&\qquad\qquad\qquad\qquad\left.+\sqrt{R_2R_3}(1-x-2xy)+\sqrt{R_1R_3}(1-3x+2xy)\right\},\\
		\Delta Q_V^{NP}&=-N_F\frac{c_{12}^V c_{23}^W c_{31}^W}{\pi^2g_V}\int_0^1dx\int_0^1dy\ \frac{x^3(1-x)y(1-y)}{\widetilde{\Lambda}},
	\end{align}
\end{subequations}
where
\begin{align}
	\widetilde{\Lambda}=-\frac{4q^2}{M_W^2}x^2y(1-y)+x^2-x(1+R_3-R_1)-(R_1-R_2)x y+R_3,
\end{align}
and $R_i=\displaystyle{\frac{m_i^2}{M_W^2}}$. 

\section{Dependence on the hypercharge in the unmixed case}\label{app:proof_Y_dependence}

The proof that the one-loop contributions to the TGC are independent of the eigenvalues of the $T^3$ operator
is as follows. For simplicity in the notation, we consider here just one copy of the multiplet. Writing the 
multiplet in terms of its $2j+1$ states, $j$ the principal quantum number, as
\begin{align}\label{eq:functionmultip}
	\Psi=\{\psi_{j,m}\}=\begin{pmatrix}
		\psi_{j,j} \\
		\psi_{j,j-1}\\
		\vdots\\
		\psi_{j,-j+1}\\
		\psi_{j,-j}
	\end{pmatrix},
\end{align}
where $m=j,j-1,\ldots,0\ (\text{or}\ \frac{1}{2},-\frac{1}{2}),\ldots,-j+1,-j$ being the 
{\it magnetic} quantum number, we first rotate to the physical gauge boson states, $W^\pm, Z^0$, and $\gamma$. Introducing the
ladder operators as usual,
\begin{align}
	T^\pm=T^1\pm i T^2,
\end{align}
together with the $T^3$ operator, we write the covariant derivative acting on the multiplet as
\begin{align}
	\mathcal{L}_G&=i\bar{\Psi}\gamma^\mu\left(\partial_\mu-i\frac{g}{\sqrt{2}}(W_\mu^+T^++W_\mu^-T^-)-i\frac{g}{c_W}(c_W^2\,T^3-s_W^2\,Y)Z_\mu-ie(T^3+Y)A_\mu\right)\Psi,
\end{align}
where $c_W=\cos\theta_W$ and $s_W=\sin\theta_W$, $\theta_W$ being the weak angle. In terms of the function multiplet
of $\Psi$, Eq. \eqref{eq:functionmultip}, we get
\begin{align}
	i\bar{\Psi}\gamma^\mu D_\mu\Psi=\sum_{m=-j}^j&\left[i\bar{\psi}_m\gamma^\mu\left(\partial_\mu-i\frac{g}{c_W}(c_W^2\,m-s_W^2\,Y)Z_\mu-ie(m+Y)A_\mu\right)\psi_m\right.\notag\\
	&\left.+\frac{g}{\sqrt{2}}\sqrt{(j+1-m)(j+m)}\, W_\mu^-\bar{\psi}_{m-1}\gamma^\mu \psi_m+\mathrm{h.c.}\right],
\end{align}
here we used the action of the ladder operators on the multiplet.

Now, we have to compute the one-loop correction to the charged TGCs coming from the
new fermions. We have to add all the possible diagrams

\begin{center}
	\includegraphics[width=\textwidth]{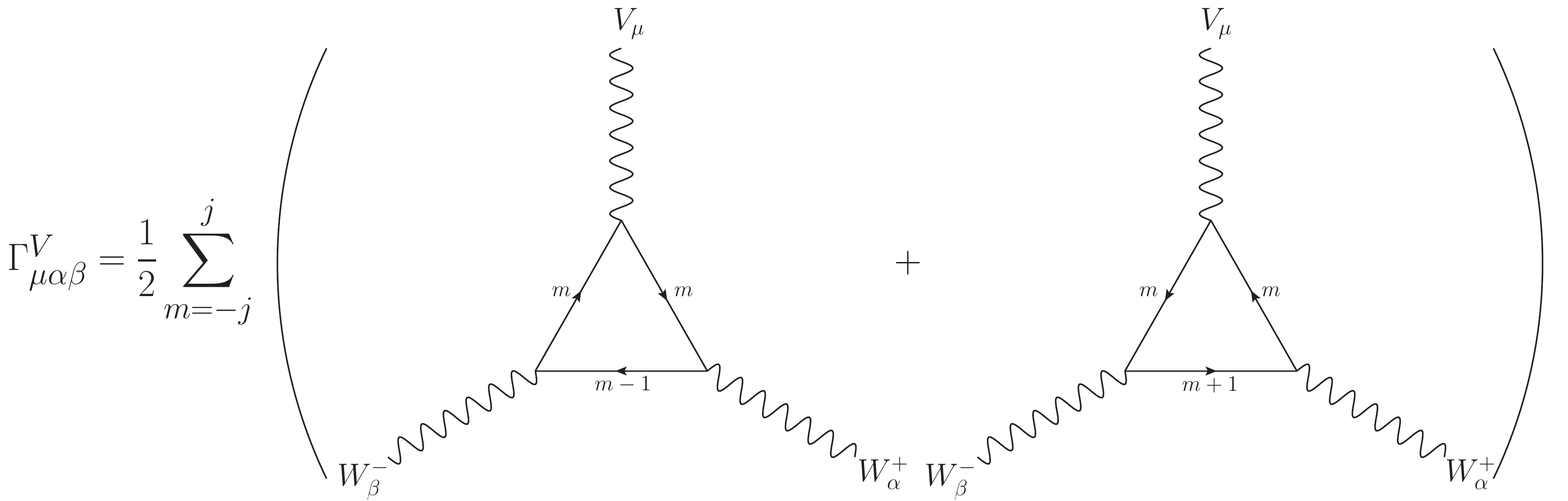}
\end{center}
  
\noindent to determine the form factors $\Delta\kappa_V^{\Psi}$ and $\Delta Q_V^{\Psi}$.
Each diagram can be written as a product of the couplings of the
fermions with the gauge bosons times a loop integral, $I_{\mu\alpha\beta}(m_m,m_m,m_{m\pm 1})$.
Therefore, the amplitude will be
\begin{align}
	\Gamma_{\mu\alpha\beta}^V=\frac{g^2}{4}\sum_{m=-j}^j &g_V^m \left[(j+1-m)(j+m)I_{\mu\alpha\beta}(m_m,m_m,m_{m-1})\right.\notag\\
	&\, +\left.(j-m)(j+m+1)I_{\mu\alpha\beta}(m_m,m_m,m_{m+1})\right],
\end{align}
where
\begin{align}
	g_V^m = 
	\begin{cases}
		e(m+Y) & \textrm{for}\ \gamma,\\
		\frac{g}{c_W}(c_W^2\,m-s_W^2\,Y) & \textrm{for}\ Z^0.
	\end{cases}
\end{align}
Since the mass of the components of the multiplet is the same, we have that 
the loop integral will depend only on the mass $m_\Psi$,
\begin{align*}
	I_{\mu\alpha\beta}(m_m,m_m,m_{m\pm 1})=I_{\mu\alpha\beta}(m_\Psi),
\end{align*}
then, the amplitude will take a simpler form,
\begin{align}
	\Gamma_{\mu\alpha\beta}^V=\frac{g^2}{2}I_{\mu\alpha\beta}(m_\Psi)\sum_{m=-j}^j g_V^m [j(j+1)-m^2].
\end{align}
Summing over the magnetic quantum number $m$,
\begin{subequations}
	\begin{align}
		\sum_{m=-j}^j [j(j+1)-m^2]&=\frac{2}{3} j (j+1) (2 j+1),\\
		\sum_{m=-j}^j m[j(j+1)-m^2]&=0,
	\end{align}
\end{subequations}
we see here that the amplitude of the one-loop correction will be proportional to the 
hypercharge,
\begin{align}
	\Gamma_{\mu\alpha\beta}^V=\frac{g^2 c_\Psi^V Y}{3}\,j(j+1)(2j+1)\,I_{\mu\alpha\beta}(m_\Psi),
\end{align}
being
\begin{align}
	c_\Psi^V = 
	\begin{cases}
		e& \textrm{for}\ \gamma,\\
		-e\, t_W& \textrm{for}\ Z^0,
	\end{cases}
\end{align}
with $t_W=\tan\theta_W$. Finally, the form factors will be computed in a standard manner. The expressions for 
$\Delta\kappa_V^{\Psi}$ and $\Delta Q_V^{\Psi}$ can be obtained from
the general expressions in Appendix~\ref{app:VFtoTGC} by taking
all the masses as identical and
\begin{align}
	c_{23}^W&=c_{13}^W=\frac{g}{\sqrt{2}}\,G_j,\\
	c_{12}^V&=c_\Psi^V\, Y,
\end{align}
where $G_j$ is the square root of the multiplet factor,
\begin{align}
	G_j=\sqrt{\frac{2}{3}\,j(j+1)(2j+1)}.
\end{align}

\bibliographystyle{JHEP}
\bibliography{TGC}

\end{document}